\documentclass[aps,prl,twocolumn,groupedaddress,reprint]{revtex4}
\usepackage{graphicx}
\usepackage{setspace}
\usepackage{amsmath}
\usepackage{amssymb}
\usepackage{color}
\usepackage{array}
\usepackage{subfigure}
\usepackage{hyperref}
\usepackage{float}
\usepackage{lipsum}
\usepackage{amsfonts}
\usepackage{mathrsfs}
\usepackage{dcolumn}
\usepackage{bm}
\usepackage{bbm}
\usepackage{graphicx}

\begin{document}
\title{Experimental Demonstration of Passive-Decoy-State Quantum-Key-Distribution with Two Independent Lasers}
\author{Shi-Hai Sun$^{1,2}$\footnote{shsun@nudt.edu.cn}, Guang-Zhao Tang$^{1,2}$, Chun-Yan Li$^{1,2}$, and Lin-Mei Liang$^{1,2,3}$ \footnote{nmliang@nudt.edu.cn}}

\affiliation{$^1$ College of Science, National University of Defense
Technology, Changsha 410073, P.R.China\\
$^2$Interdisciplinary Center for Quantum Information, National University of Defense Technology, Changsha 410073, P.R.China\\
$^3$State Key Laboratory of High Performance Computing, National University of Defense Technology, Changsha 410073, P.R.China}

\date{\today}
\begin{abstract}
Decoy state method could effectively enhance the performance of quantum key distribution (QKD) with practical phase randomized weak coherent source. Although active modulation of the source intensity is effective and has been implemented in many experiments, passive preparation of decoy states is also an important addition to the family of decoy state QKD protocols. In this paper, following the theory of Curty \emph{et al.} [PRA, 81, 022310 (2010)], we experimentally demonstrate the phase-encoding passive-decoy-state QKD with only linear optical setups and threshold single photon detectors. In our experiment, two homemade independent pulsed lasers, with visibility of Hong-Ou-Mandel interference $0.53(\pm 0.003)$, have been implemented and used to passively generate the different decoy states. Finally, secret key rate $1.5\times 10^{-5}$/pulse is obtained with 10km commercial fiber between Alice and Bob.
\end{abstract}


\maketitle
\emph{Introduction-} Quantum key distribution (QKD) \cite{BB84} admits two parties to share an unconditional secret key, which is guaranteed by the basic principle of quantum mechanics. Until now, high-speed and long-distance QKD have been implemented both in the laboratory and in the field \cite{Yuan08,Wang12,Chen10,Tang14}. However, due to the imperfection of practical electrical and optical setups, potential loopholes in practical QKD systems could be exploited by an eavesdropper (Eve) \cite{Zhao08,Lydersen10,Xu10,Jain11,Sun12,Bugge14} to spy the final secret key. Luckily, some countermeasures could be used to cover the gap between the theory and the practice. One of the most famous cat-and-mouse games is the photon-number-splitter (PNS) attack \cite{Brassard00} and decoy state method \cite{Hwang03,Wang05,Lo05}. Due to the unavailability of the single photon source, phase randomized weak coherent source (PR-WCS) is always used in many practical QKD systems. However, the photon number distribution of the PR-WCS is Poisson. And the multi-photon pulses will leave potential loophole for Eve to perform the PNS attack, in which Eve blocks all single photon pulses and splits one photon from other $n>1$ photon pulses. Then the secret key rate will be dramatically decreased and the maximal secret distance will be limited within tens of kilometers. In order to defeat such loophole, decoy state method was proposed to strictly estimate the yield and error rate of single photon pulses.

As a necessary technology to effectively enhance the performance of QKD with PR-WCS, decoy state QKD has been implemented in many experiments \cite{Rosenberg07,Schimtt07,Peng07,Dixon08}. In most of these experiments, Alice actively modulates the intensity of source to generate different decoy states. Although active intensity modulation with intensity modulator is effective, passive preparation of decoy states is also desirable. The parametric down-conversion (PDC) source, in which the photon number of two output modes is strongly correlated, has been considered as a built-in decoy state by measuring one output mode of the PDC. Passive decoy state based on PDC has been proposed by many groups \cite{Mauerer07,Adachi07,MaXF08}, and recently experimentally demonstrated \cite{SunQC14}. However PDC source is difficult in experiments, which may weaken its performance in many practical applications. In order to overcome the gap, passive decoy state with practical PR-WCS has been proposed \cite{Curty09,Curty10}, in which different non-Poisson signal pulses could be generated with only linear optical elements and threshold photon detectors. Then it was partially demonstrated in experiment \cite{Zhang10,Zhang12}. In Ref.\cite{Zhang10}, the authors only implemented the intrinsic-stable non-Poisson light source but not a complete QKD protocol. In Ref.\cite{Zhang12}, the authors implemented a modified passive decoy state protocol with the PR-WCS and the Faraday-Michelson QKD system \cite{Mo05}. In these experiments (both Ref.\cite{Zhang10} and Ref.\cite{Zhang12}), Alice uses one PR-WCS and one unbalanced interference (unbalanced Mach-Zehnder interferometer or unbalanced Faraday-Michelson interferometer) to passively generate different non-Poison lights. Although only one laser is required, there are some inherent drawbacks for such system. First, the repetition of the system must match with the difference of two arms of the unbalanced interference. Second, there may exist phase correlation between adjacent pulses \cite{Kobayashi14}, which will affect the photon number distribution of different decoy states and then compromise the security of QKD protocol.

In this paper we experimentally demonstrate the phase-encoding passive decoy state QKD with two independent lasers. The visibility of Hong-Ou-Mandel (HOM) interference for the homemade independent lasers reaches $0.53(\pm 0.003)$. Then different decoy states could be passively generated based on the respondence of Alice's threshold single photon detector (SPD). Finally, the secret key rate about $1.5\times10^{-5}/$pulse is obtained with about 10km commercial fiber between Alice and Bob. Our results show that the passive decoy state method with practical PR-WCS is possible and has potential applications in practices.

\begin{figure}
\subfigure{
\label{fig:mini:subfig:a} 
\begin{minipage}[b]{0.25\textwidth}
\centering
\includegraphics[width=2in]{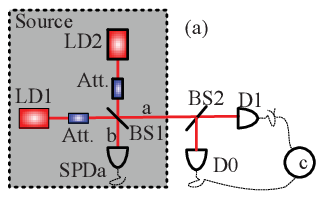}
\end{minipage}}%
\subfigure{
\label{fig:mini:subfig:b} 
\begin{minipage}[b]{0.25\textwidth}
\centering
\includegraphics[width=1.5in]{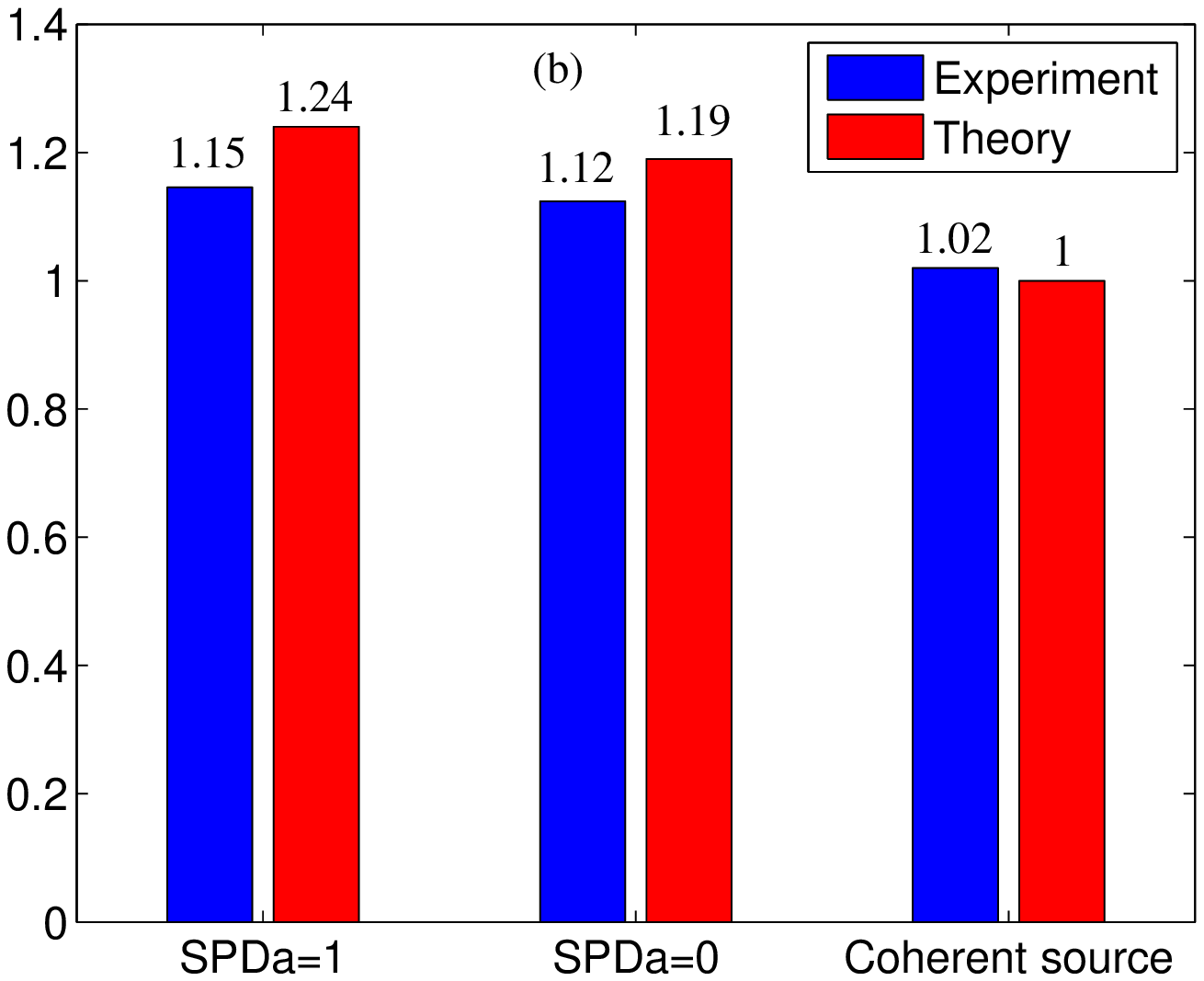}
\end{minipage}}
\caption{(Color online) Non-Poisson source generation and HBT experiment.
(a) shows the setups for the non-Poisson source generation and the scheme for HBT experiments. (b) shows the measured $g^{2}$ for the generated
non-Poisson source. Here $SPDa=1$ (and $SPDa=0$)means the SPDa clicks (and non-clicks). The standard deviation of the experimental results are 0.06, 0.03 and 0.03 for $SPDa=1$, $SPDa=0$ and the coherent source, respectively. For each case, the blue bar (right) and the red bar (left) show the experimental results and theoretical values, respectively.
The accumulated time for each bar is 600s.}
\label{fig:HBT_exp} 
\end{figure}

\emph{Passive decoy state-} We first briefly review the passive decoy state method following the theory of Ref.\cite{Curty10}. The basic setup of the passive decoy state method is shown in Fig.\ref{fig:mini:subfig:a} (a). Two independent lasers (noted as LD1 and LD2 respectively) with different intensities interfere at a beam splitter (BS1). The transmittance of BS1 is noted as \emph{t}. Alice measures the light in one mode of the BS (mode \emph{b}) with a SPD (noted as SPDa). When the SPD clicks, Alice notes the pulses in mode \emph{a} of the BS1 as signal state, otherwise, she notes them as decoy state. The density matrixes of LD1 and LD2 are given by
\begin{equation}
\begin{split}
\rho=e^{-\mu_1}\sum_{n=0}^\infty \frac{\mu_1^n}{n!}|n\rangle\langle n|,\\
\sigma=e^{-\mu_2}\sum_{n=0}^\infty \frac{\mu_2^n}{n!}|n\rangle\langle n|,
\end{split}
\end{equation}
here $\mu_1$ and $\mu_2$ are the average intensities of LD1 and LD2, respectively. The joint probability that \emph{n} photons in mode \emph{a} of BS1 and \emph{m} photons in mode \emph{b} of BS1 can be written as
\begin{equation}
P_{n,m}=\frac{\nu^{n+m}e^{-\nu}}{2\pi n!m!}\int_0^{2\pi}\gamma^n(1-\gamma)^m d\theta,
\end{equation}
where
\begin{equation}
\begin{split}
\nu=\mu_1+\mu_2,\\
\gamma=\frac{\mu_1 t+\mu_2(1-t)+\xi \cos(\theta)}{\nu},\\
\xi=2\sqrt{\mu_1\mu_2(1-t)t}.
\end{split}
\end{equation}

Then the joint probability that \emph{n} photons in mode \emph{a} of BS1 and no click in SPDa, and the joint probability that \emph{n} photons in mode \emph{a} of BS1 and SPDa clicks are given by
\begin{equation}\label{Eq_pro}
\begin{split}
P_n^{nc}&=(1-\epsilon)\sum_{m=0}^{\infty}(1-\eta_d)^m P_{n,m},\\
P_n^{c}&=\sum_{m=0}^{\infty}P_{n,m}-P_n^{nc}\equiv P_n^t-P_n^{nc}.
\end{split}
\end{equation}
Here the subscript $nc$ (or $c$) means the SPD of Alice doesn't click (or clicks). $\epsilon$ and $\eta_d$ are the dark count rate and efficiency of SPDa. It is easy to check that the probability distributions of $P_n^{nc}$ and $P_n^{c}$ are non-Poisson.

Then Alice and Bob could estimate the secret key rate by combining the the GLLP formula \cite{GLLP04} and the idea of decoy state method, which is given by Ref.\cite{Curty10}
\begin{equation}
R\geq \sum_l \max\{R^l, 0\},
\end{equation}
where $l\in\{c, nc\}$ which means SPDa clicks or doesn't click, and
\begin{equation}
R^l\geq q\{-Q^l f(E^l) H(E^l)+(P_1^l Y_1^L+P_0^l Y_0^L)[1-H(e_1^U)]\}.
\end{equation}
Here $q$ is the efficiency of the QKD protocol ($q=1/2$ for BB84 protocol \cite{BB84}); $f(E^l)$ is the efficiency of the error correction protocol; $Q^l$ ($E^l$) is the total gain( error rate); $Y_1^L$ ( $e_1^U$) is the lower bound of yield (upper bound of error rate) of the single photon pules. $Y_0^L$ is the lower bound of dark count rate of Bob's SPD. $P_1^l$ ($P_0^l$) is the probability of single photon pules (vacuum pulse).

Finally, according to theoretical analysis of Ref.\cite{Curty10}, the lower bound of yield and the upper bound of the error rate for the single photon pulse are given by
\begin{subequations}
\begin{equation}
\begin{split}
P_1^l Y_1^L+P_0^l Y_0^L= &\max \{ \frac{P_1^l(P_2^t Q^{nc}-P_2^{nc}Q^t)}{P_2^tP_1^{nc}-P_2^{nc}P_1^t}\\
& +[P_0^l-P_1^l\frac{P_2^tP_0^{nc}-P_2^{nc}P_0^t}{P_2^tP_1^{nc}-P_2^{nc}P_1^t}]Y_0^U, 0\},
\end{split}
\end{equation}
\begin{equation}
\begin{split}
e_1^U=&\min\{\frac{E^c Q^c-P_0^c Y_0^Le_0}{P_1^c Y_1^L}, \frac{E^{nc} Q^{nc}-P_0^{nc} Y_0^Le_0}{P_1^{cn}Y_1^L},\\
&\frac{P_0^{nc}E^t Q^t-P_0^t E^{nc}Q^{nc}}{(P_1^tP_0^{nc}-P_1^{nc}P_0^t)Y_1^L}\},
\end{split}
\end{equation}
\end{subequations}
where $e_0=1/2$ is the error rate of background, and
\begin{equation}
\begin{split}
Y_0^U=\min\{\frac{E^cQ^c}{P_0^ce_0}, \frac{E^{nc}Q^{nc}}{P_0^{nc}e_0}\},\\
Y_0^L=\max\{\frac{P_1^tQ^{nc}-P_1^{nc}Q^t}{P_1^tP_0^{nc}-P_1^{nc}P_0^t},0\},\\
Q^t=Q^c+Q^{nc},\\
Q^tE^t=Q^cE^c+Q^{nc}E^{nc}.
\end{split}
\end{equation}

\emph{Experiment-} In the passive decoy state scheme, non-Poisson source is generated with two PR-WCS. The generation setups of the non-Poisson is shown in Fig.\ref{fig:HBT_exp}, in which two PR-WCS interfere at a beam splitter (BS1). One mode of BS1 (noted as mode \emph{b}) is measured with a SPD (noted as SPDa in experiment since the detector belongs to Alice), and the other mode of BS1 (note as mode \emph{a}) is used as signal state or decoy state depending the click of the SPDa. Note that it is also possible to implement the passive decoy state generation with one laser and one unbalanced Mach-Zehnder interferometer (UMZI). But there are some drawbacks in such scheme. For example, the phase of different pulse may not totally independent \cite{Kobayashi14}. Then security should be reevaluated to remove the relationship of phase within different pulses. Thus, in our experiment, two indistinguishable independent lasers are implemented to passively generate different decoy states. Furthermore, we note that although two weak coherent lights are used to passively generate the signal state and decoy state in our experiment, it is still possible to generate the non-Poisson source with strong coherent light combining with classical threshold detector \cite{Curty10}.

In our implementation, interference between two lasers is required, which could be characterized by the visibility of HOM interference. The theoretical value is 0.5 for the coherent source with random phase, however the measured visibility is about $0.53\pm0.003$ in our experiment. Here we give some discussion about the imperfection of the HOM interference.

In order to ensure that the pulses from LD1 and LD2 could interfere at the BS1, the photons should be indistinguishable in polarization, spectrum, time. Any mismatch in these dimensions will affect the photon number distribution of different decoy states, and then worsen the performance of the passive decoy state QKD protocol. The polarization is automatically matched using polarization maintain fiber from the laser diodes to the BS1 in our experiment. Although, strictly speaking, the axes of the fiber may mismatch in practical experiment, the error introduced by it is small.
The butterfly DFB laser diode is used in our experiment, whose 3dB width of spectrum is about 60pm. By carefully modulating the temperature of the laser diode, the difference of the center wavelength between LD1 and LD2 can be set small enough. In our experiment, the center wavelength of laser diodes is set as 1559nm with difference less than 10pm, which is less than the spectrum of the laser diodes.

The major difficult for the HOM interference between two independent lasers is the arriving time of the photons. In our experiment, in order to increase the visibility of HOM interference, a homemade electrical delay with step 10ps is used to adjust the trigger time of LD1 and LD2. However, we should note that although the precision of our delay device is high enough, the temporal mode of optical pulse is still the main imperfection for the visibility of HOM interference due to the time jilter of electrical devices. In our experiment, the time jilter is about 100ps, thus the width of optical pulses from LD1 and LD2 is set as 2ns to reduce the effect of time jilter. Broad pulses are also used in many MDI-QKD experiments \cite{Ferreira13,Tang-Yin14}. It seems that it will be a bottleneck for the implementation of the passive decoy state QKD or MDIQKD with future high speed operation. However, recently, MDI-QKD based on two independent lasers with repetition rate 1GHz has been demonstrated in experiment \cite{Camandar16}, in which the width of pulse is 35ps. Thus we think high speed QKD with passively decoy state generation is still possible by improving the performance of the devices.

To evaluate the non-Poisson statistics of the two kinds of pulses, signal state for SPDa click and decoy state for SPDa non-click, a HBT experiment is performed with two SPDs (ID201, Idquantique). Here, we use the correlation function of optical pulses, $g^{(2)}$, to characterize the non-Poisson statistics of the pules. In our experiment, the average intensity of LD1 (and LD2) is set as 0.64 (and 0.08). Then the theoretical predictions of $g^{(2)}$ for signal state (SPDa click) and decoy state (SPDa non-click) are 1.24 and 1.19, respectively. With the experimental setups of Fig.\ref{fig:HBT_exp}, the measured $g^{(2)}$ is 1.15 with standard deviation 0.06 for the pulses that SPDa clicks, and 1.12 with standard deviation 0.03 for the pulses that SPDa non-clicks. All the results are shown in Fig.\ref{fig:HBT_exp}(b). Here we also measure $g^{(2)}$ for the coherent state. The measured $g^{(2)}$ is 1.02 with standard deviation 0.03, which is very close the theoretical prediction of 1.

\begin{figure}
\scalebox{1}{\includegraphics[width=\columnwidth]{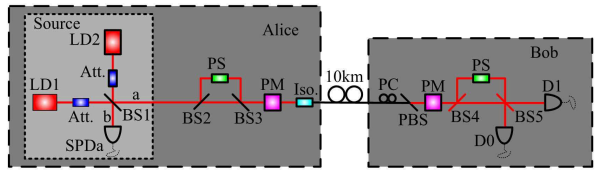}}
\caption{\label{fig:scheme}(Color online) Schematic setup for QKD protocol. LD1 and LD2 are distributed feedback laser diodes. Att. is attenuator used to modulate the intensity of signal pulses from LD1 and LD2. BS: beam splitter; PS: phase shift; PM: phase modulator; Iso.: isolator; PC: polarization controller; PBS: polarization beam splitter. SPDa is the single photon detector of Alice, which is used to determine that the pulse is signal state or decoy state depending on the click of it. D0 and D1 are single photon detectors of Bob. The red lines are polarization maintain fiber. Alice and Bob are connected with about 10km commercial fiber.}
\end{figure}

Then with the non-poisson source given above, we perform the QKD based on BB84 protocol. The setups are shown in Fig.\ref{fig:scheme}. The pulses on mode \emph{a} pass through the UMZI, in which a phase shift (PS) is used to compensate the phase between the long arm and short arm. The encoding phase of Alice is modulated on the pulses that pass through the short-arm of the UMZI with a phase modulator (PM). At the same time, in order to remove the Trojan-horse attack \cite{Gisin06}, an isolator is used to stop any light to be injected into Alice's zone form channel. When the pulses arrive at Bob's zone, the polarization is controlled by combing a polarization controller (PC) and a polarization beam splitter (PBS). The decoded phase of Bob is modulated on the pulsed that pass through the long arm of Alice's UMZI. Then Bob uses a UMZI and two SPDs (D0 and D1) to measure Alice's information.

The repetition frequency of our system is 2.5MHz, which is limited by the maximal repetition of Bob's SPD (iD201, Idquantique). The intensities of LD1 and LD2 are set as about 0.64 and 0.08, respectively. And the transmittance of BS1 is 0.5 in our experiment. Then pulses on mode \emph{b} of BS1 are detected by Alice's SPD, whose dark-count rate is about $1.2\times 10^{-5}$/pulse with a gate width of 2.5ns and an efficiency of 10\%. Then final secret key rate is estimated. All the experimental results are listed in Table \ref{tab:Table 1}. Note that strictly speaking, the statistical fluctuation of the intensity of LD1 and LD2 should be taken into in the estimation of final key rate. However, as a proof-of-principle proof, we assume the intensities of LD1 and LD2 are stable in this paper. By controlling the temperature of laser diodes, the intensities of LD1 and LD2 are very stable. In fact, the measured standard deviations in one hour for LD1 and LD2 are 0.005 and 0.001, respectively. The decrease of key rate caused by the intensity fluctuation of LD1 and LD2 could be ignored \cite{Yuan14}.

\begin{table}
\caption{\label{tab:Table 1} Experimental results of our experiment. Here, $N$ is the length of collected data; $t$ is the transmittance of BS1; $\mu_1$ ($\mu_2$) is the average photon number of LD1 (LD2); $E^c$ ($E^{nc}$) is the total error rate given that Alice's SPD clicks ( does not click);  $Q^c$ ($Q^{nc}$) is the total gain given that Alice's SPD click (does not click); $R$ is the final secret key rate. $f(E)=1.22$. The accumulated time for the experiment that measures the stability of lasers is one hour.}
\tabcolsep0.02in
\doublerulesep2pt
\begin{tabular}{cc|cc}
\hline\hline
Parameter &Result &Parameter & Result \\
\hline
$\mu_1$ & 0.64($\pm0.005$) & $\mu_2$ & 0.08($\pm0.001$) \\
$E^c$ & 6.13($\pm3.42$)\%  & $E^{nc}$ & 5.55 ($\pm 0.52$)\% \\
$Q^c$ & 2.54($\pm0.35$)$ \times 10^{-6}$ & $Q^{nc}$ & 8.18($\pm0.21$)$\times10^{-5}$  \\
R & $1.50\times 10^{-5}$ & &\\
\hline\hline
\end{tabular}
\end{table}

In our experiment, the estimated final secret key rate is about $1.50\times 10^{-5}$/pulse with only 10km commercial fiber between Alice and Bob. The secret key rate is much lower than the active decoy state QKD experiment. The main reason is that, in passive decoy state method, the intensity of Alice's pulses should be attenuated to weak light before the BS1, but not at the end port of Alice (after the Iso.). Thus the loss of Alice's optical setups should be taken into account in the passive decoy state method (generally speaking, the loss of Alice's optical setups could be ignored in the active decoy state method). This is the main reason that the key rate of our experiment is much lower than the active decoy state method. It seems that it is a major disadvantage for the passive decoy state method. However, we think this drawback could be improved to enhance the performance of the passive decoy state method. First, the loss of Alice's optical setups is about 9dB in our experiment, which could be reduced by using low loss optical devices. Second, as a proof-of-principle experiment, the parameters are not optimized in our experiment, thus the final key rate could be increased by optimizing all the experimental parameters. Third, the legitimate parties could use the strong coherent light scheme to replace the weak coherent light scheme \cite{Curty10}. Then the loss of Alice's setups could be compensated.Thus, the passive decoy state still has potential advantages and could be applied in future.

\emph{Conclusion-}
In this paper, the phase-encoding passive decoy state QKD has been experimentally implemented with only linear optical setups and threshold SPDs. The different decoy states could be generated based on the HOM interference with two homemade independent pulsed lasers. The visibility of HOM interference reaches $0.53(\pm 0.003)$ by modulating the central wavelength with temperature controller and the arriving time with electrical delay chip. The final secret key rate $1.50\times 10^{-5}/$pulse is obtained. Our experiments clearly show that the passive decoy state method with practical PR-WCS is also desirable.

\emph{Acknowledgments-} The authors thank M.S.Jiang for helpful discussions of the electrical driver of laser diodes. This work is supported by the National Natural Science Foundation of China, Grant No. 11304391, 11674397.


\end{document}